\RequirePackage{snapshot}
\documentclass[journal,11pt,twoside]{IEEEtranTCOM}
\usepackage{amsfonts,amsmath,amstext,verbatim}
\usepackage{amssymb}%
\usepackage{setspace}
\newcommand{\resetcounters}{\setcounter{equation}{0}}
\usepackage{amsfonts,amsmath,amstext,verbatim}
\usepackage{amssymb}%
\usepackage{setspace}
\usepackage{ifpdf}

\newcommand{\squishlist}{
 \begin{list}{{\bf --}}
  { \setlength{\itemsep}{0pt}
     \setlength{\parsep}{1pt}
     \setlength{\topsep}{0pt}
     \setlength{\partopsep}{0pt}
     \setlength{\leftmargin}{0.5em}
     \setlength{\labelwidth}{1em}
     \setlength{\labelsep}{0.2em} } }

\newcommand{\squishlisttwo}{
 \begin{list}{$\bullet$}
  { \setlength{\itemsep}{0pt}
     \setlength{\parsep}{0pt}
    \setlength{\topsep}{0pt}
    \setlength{\partopsep}{0pt}
    \setlength{\leftmargin}{2em}
    \setlength{\labelwidth}{1.5em}
    \setlength{\labelsep}{0.5em} } }

\newcommand{\squishend}{
  \end{list}  }

\renewenvironment{abstract}{%
\small\bfseries\textit{Abstract}:  }

\ifCLASSINFOpdf
\usepackage[pdftex]{graphicx}
\usepackage[pdftex,colorlinks,%naturalnames,%
           bookmarks=true,%
           pdftitle=Interference\ and\ SINR\ coverage\ in\ spatial\ non-slotted Aloha\ networks,%
           pdfauthor=B.\ Blaszczyszyn%
           \ -\  P.\ Muhlethaler,]{hyperref} 
\usepackage[numbers,sort&compress]{natbib}
\usepackage{hypernat}
\else

\usepackage[dvips]{graphicx} 
\usepackage[dvips,colorlinks,%naturalnames,%
           bookmarks=true,%
           pdftitle=Interference\ and\ SINR\ coverage\ in\ spatial\ non-slotted Aloha\ networks,%
           pdfauthor=B.\ Blaszczyszyn%
           \ -\  P.\ Muhlethaler,]{hyperref} 
\usepackage[numbers,sort&compress]{natbib}
\fi

\hypersetup{
   bookmarksnumbered,
   pdfstartview={FitH},
   citecolor={blue},
   linkcolor={red},
   urlcolor=[rgb]{0,0.55,0},%{green},
   pdfpagemode={UseOutlines}}

\newcommand{\md}{\text{\rm d}}

\newcommand{\E}{{\mathbf E}}
\newcommand{\Pro}{{\mathbf P}}
\newcommand{\ind}{1\hspace{-0.30em}{\mbox{I}}}

\renewcommand{\theequation}{\arabic{section}.\arabic{equation}}

\newtheorem{Th}{Theorem}[section]
\newtheorem{prop}[Th]{Proposition}
\newenvironment{Prop}{\bf\begin{prop}\rm\em}{\end{prop}} % proposition
\newtheorem{res}[Th]{Result}
\newenvironment{Res}{\bf\begin{res}\rm\em}{\end{res}} % Result
\newtheorem{fact}[Th]{Fact}
\newtheorem{Rem}[Th]{Remark}
\newenvironment{rem}{\bf\begin{Rem}\rm}{\end{Rem}} % Numbered remark

\newcommand{\bfF}{\mathbf{F}}

\newcommand{\calR}{{\mathcal{R}}}
\newcommand{\calL}{{\mathcal{L}}}

\newcommand{\bbZ}{\Bbb{Z}}
\newcommand{\bbR}{\Bbb{R}}

\makeatother  \renewenvironment{abstract}{%
  \small\bfseries\textit{Abstract}:  }

\makeatother  \renewenvironment{abstract}{%
  \small\bfseries\textit{Abstract}:  }

\begin{document}
%\title{Stochastic Analysis of a Linear Vehicular Ad-hoc NETwork
%(VANET)}
%\title{Modeling and analysis of linear Vehicular\ Ad-hoc 
% NETworks with Aloha}
\title{Interference and SINR coverage in spatial non-slotted Aloha networks 
%%Interference in Pure Spatial Aloha Ad-hoc Networks 
}
%% \author{Bart{\l }omiej B{\l }aszczyszyn\authorrefmark{1}, Paul
%%   M{\"u}hlethaler\authorrefmark{2} and Yasser Toor\authorrefmark{2}
%% \thanks{\authorrefmark{1}INRIA-ENS,
%% 23 Avenue d'Italie  75214 Paris, France
%% e-mail: Bartek.Blaszczyszyn@ens.fr}
%% \thanks{\authorrefmark{2}INRIA  Rocquencourt, Le Chesnay, FRANCE,
%% e-mail: \{Paul.Muhlethaler,Yasser.Toor\}@inria.fr}}
\author{
\begin{tabular}{c c c c }
Bar{\l}omiej B{\l}aszczyszyn & Paul M{\"u}hlethaler  \\
Inria/ENS &  Inria  Rocquencourt \\
Paris FRANCE &  Le Chesnay FRANCE \\
Bartek.Blaszczyszyn@ens.fr & Paul.Muhlethaler@inria.fr \\
\end{tabular}
\vspace{-4ex}
}

\maketitle
\thispagestyle{empty} \pagestyle{empty}

\begin{abstract}
In this paper we propose two analytically tractable
stochastic-geometric models of interference in ad-hoc networks using  
pure (non-slotted) Aloha as the medium access. In contrast the slotted model, the interference in pure Aloha 
may vary during the transmission of a tagged packet. 
We develop closed
form expressions for the Laplace transform of the empirical average 
of the  interference experienced 
during the transmission of a typical packet.  
Both models  assume a power-law path-loss function with arbitrarily
distributed fading and feature configurations of 
transmitters randomly located in
the Euclidean plane according to a Poisson point process. Depending on the
model, these configurations vary over time or are static.
We apply our analysis  of the interference 
to study the Signal-to-Interference-and-Noise Ratio (SINR)   
outage probability for a typical transmission in pure Aloha.
The results are used to compare  the performance of non-slotted Aloha
to the slotted one, which has almost exclusively been previously 
studied in the same context of mobile
ad-hoc networks.

\end{abstract}


\vskip -0.7 cm
\section{Introduction}
\label{s.I}

Aloha is one of the simplest multiple access communication protocols. The 
main idea is that sources, independently of each other,
 transmit  packets and back-off for  
random times before the next transmission.
A classical model of Aloha assumes that when two or more transmissions overlap in 
time then neither of them is successful (a collision occurs). This 
simple  model is not adequate in the wireless context. Indeed, the 
geometry of nodes in 
the network may allow some simultaneous transmissions to be successful due to 
the capture (or spatial reuse) effect. 
Since 1988 and the seminal paper~\cite{verdu88}, new models have been developed to describe 
Aloha networks with spatial reuse and capture effect. The model
proposed in~\cite{BBM06IT}, which inspired our work, falls into this category. 
To the best of our knowledge, almost all of these studies use slotted Aloha, 
except~\cite{BBPMnsaloha09}, which we revisit in this paper. In Section~\ref{sec:model}, we consider a wireless ad-hoc network 
modeled by homogeneous Poisson point process. 
We assume power-law decay of the signal in the path-loss model as 
well as a general distribution of the fading. 
We consider two models for non-slotted Aloha: the {\em Poisson rain} model 
with dynamic node activation and 
the {\em Poisson renewal} model with a static positioning of nodes. 
In both of these models, we characterize 
the Laplace transform of the distribution of the empirical average of the interference 
received during the reception of a typical packet, see Section~\ref{sec:IA}. The 
expressions are particularly simple for 
the {\em Poisson rain} model.
Next in Section~\ref{sec_sinr}, we use this characterization of the Laplace transform of the interference 
to express 
the probability of a successful transmission in pure Aloha. 
Finally, we optimize and compare 
space-time and energy efficiency of pure Aloha with the previously studied performance
of slotted Aloha (Subsection~\ref{subsec:spacetime}).
We show that  the {\em Poisson rain} and 
the {\em Poisson renewal} models provide comparable results and that 
 the {\em Poisson renewal} model converges towards the {\em Poisson rain} 
model when the node density increases.

Comparison of slotted and non-slotted Aloha is a classical subject. In 
a simple but widely used model where the aggregate packet transmission 
process follows a Poisson distribution (i.e the spatial 
reuse is not taken into account) pure ({\em non-slotted}) Aloha   
can on average attain the fraction $1/(2e)\approx18.4\%$  of
successful transmissions, when the scheme is optimized  by tuning 
the  mean back-off time. 
It can be shown that this performance can be multiplied
by~2 in {\em slotted-Aloha}, when all the nodes are synchronized and
can send packets only at the beginnings of some universal time slots.
In this paper we show that taking into account the spatial distribution 
of transmitters makes the comparison of both types of Aloha more subtle. 
Specifically, we observe that when the path-loss exponent is not
very strong both Aloha protocols, when appropriately optimized, exhibit a
similar performance (see Figure~\ref{fig.cpdebitop_bvar1}). Hence the extra complexity required by slotted Aloha may be questionable. For stronger 
path-loss, slotted  Aloha
offers a gain (due to synchronisation), which is however much smaller than 
that predicted by the classical (geometry-less) models of Aloha,
which do not take into account the spatial reuse effect but 
assume collisions for all simultaneously transmitted packets.
Moreover,  we observe in  all path-loss scenarios that 
both slotted and pure 
Aloha schemes exhibit the same energy
efficiency (see Subsection~\ref{subsect:enereff}). 

{\color{black} The results presented in this paper allow the 
performance 
of slotted and non-slotted Aloha with spatial reuse to be evaluated 
directly (or indirectly via numerical computations). 
The tools introduced in this paper can handle  
various fading scenarios.   
Although the main-stream protocols currently used are 
based on the Carrier Sense Multiple Access 
(CSMA) scheme, 
we know that these protocols do not always work satisfactorily. For instance, 
we know that the hidden-terminal problem is a very serious issue for CSMA, 
which, in such a situation, behaves at most no better than an Aloha protocol. 
Tuning the 
carrier sense threshold of CSMA is also very difficult when the node density 
of the network varies greatly. In these situations Aloha protocols 
can be a perfectly acceptable alternative to CSMA protocols. 
In all our Aloha models the 
back-off is taken into account via the medium access probability or explicit 
back-off times. }

%With Rayleigh fading we can compute the capture probability with 
%a Signal over Interference Ratio 
%(SIR) model for both slotted and pure Aloha, we obtain . 
%compute the Laplace 
%transform of the mean interference with closed formulas.   
%Another goal of this paper is to compare slotted and non-slotted 
%Aloha. In the Poisson rain model of non-slotted Aloha, we can precisely 
%compare the Laplace transform of the mean interference for 
%slotted Aloha and non-slotted Aloha. With Rayleigh fading we can 
%compute the capture probability with a Signal over Interference Ratio 
%(SIR) model. This allows the throughput of slotted Aloha and non-slotted 
%Aloha to be compared. 

\subsection{Related Work}
\label{s.RW}
The shot noise process formed from a sum of response functions of a
Poisson process and identically distributed random variables
was already studied back in the 60s, cf~\cite{gilbert1960amplitude}.
The first attempts to study interference using the Laplace
transforms of the shot noise can be found in~\cite{zorzi1994outage,sousa1990optimum}.
In~\cite{lowen1990power}
the shot noise created in a Poisson process with a power-law path-loss function was 
studied and observed to have a stable distribution. 

In \cite{BBM06IT} the authors 
introduce what is today called the Poisson bipolar model of an ad-hoc
network and using the Laplace transform of the interference they calculate the 
SINR capture probability for a typical packet transmission in the slotted Aloha MAC. 
This approach has been next extended. 

%% In particular in~\cite{andrews05} the authors compute upper and lower bounds 
%% on the transmission capacity of wireless ad-hoc spread spectrum networks. 
%% Using these bounds, they show that frequency hopping systems obtain
%% higher transmission capacity than direct sequence systems in the 
%% order of $M^{1-2/\beta}$, where $M$ is the spreading factor and $\beta$ is 
%% the path loss exponent.

In~\cite{Haenggi2009} the outage probability of a packet in a 
Signal-over-Interference-Ratio model is extensively investigated. This 
paper considers many situations: Poisson point process and 
deterministic node placement. Various kinds of fading are studied. 
The study also varies the medium access scheme considering 
Aloha and Time Division Multiple Access (TDMA). The author computes the 
outage probability of a packet in all these situations. 
In~\cite{JSAC} a channel-fading opportunistic Aloha is considered. 

%% Although most of the studies use Aloha for the access scheme, a few of them 
%% such as \cite{ngu07} adopt CSMA. Studying 
%% the interference in CSMA networks is more difficult because the 
%% pattern of the simultaneously transmitting nodes is no longer a Poisson process
%% even if the initial node location is assumed to be Poisson. A natural model 
%% is a Matern-like hard-core (or soft-core in the case of fading) point process.   

Most research papers on Aloha and its performance concern 
slotted Aloha. Very few papers study pure (non-slotted) Aloha. The early
papers on pure Aloha such as~\cite{rob75} do not consider spatial reuse.
The present paper, inspired by the original approach of~\cite{BBM06IT}, revisits and extends the study of pure Aloha
proposed in~\cite{BBPMnsaloha09}.

The remaining part of this paper is organized as follows. Section~\ref{sec:model} describes 
the two models for pure Aloha: the {\em Poisson rain } and 
the {\em Poisson renewal} models. Section~\ref{sec:IA} analyzes the interference in our Aloha 
networks by computing the Laplace transform of the mean interference 
for pure Aloha in  the {\em Poisson rain } and 
the {\em Poisson renewal} models for a general fading. In 
Section~\ref{sec_sinr}, we study the SINR coverage probability using the 
Laplace transform of the empirically averaged interference. 
We can thus optimize 
pure Aloha and compare the performance with the slotted version 
of the protocol. Section~\ref{sec:comp} contains the simulation results which validate 
our models and also provides numerical comparisons of slotted and pure Aloha. 
Section~\ref{sec:con} concludes the paper. The appendix provides the computation 
of  the Laplace transform of the mean interference with pure Aloha in the 
{\em Poisson renewal} model. This computation is somewhat technical, which is  
why we include it in the Appendix. 
%%Haenggi Andrews + ad hoc 1 ou 2 papiers

\section{Network and Aloha model }
\label{sec:model}

In this section we present our models of 
non-slotted Aloha for wireless ad-hoc
networks. To facilitate future comparisons, we also recall the basic spatial
slotted Aloha model.

\subsection{Location of Nodes --- The Spatial Poisson Bipolar Network
  Model} 
\label{ss.Bipolar}
We consider a {\em Poisson bipolar network model}
in which each point of the Poisson pattern 
represents a node of a Mobile Ad hoc NETwork (MANET) and is hence a potential transmitter.
Each node %has an infinite backlog of packets to transmit and 
has  an associated receiver located at distance $r$. This receiver is 
not part of the Poisson pattern of points.
Using the formalism of the theory of point
processes, we will say that a snapshot of the MANET 
can be represented by an independently marked Poisson point process (P.p.p)
$\widetilde\Phi=\{(X_i,y_i)\}$,
where the {\em locations of nodes}  $\Phi=\{X_i\}$ form  a homogeneous P.p.p. on
the plane, with an intensity of $\lambda$ nodes per unit of space,
and where the mark $y_i$
denotes the location of the receiver for node $X_i$.
We assume here that no two transmitters have the same receiver
and that, given $\Phi$,  the vectors $\{X_i-y_i\}$ are i.i.d  with
$|X_i-y_i|=r$. {\color{black} 
In this paper $r$ is constant however, in principle, 
the results obtained can be 
extended by integrating the final formulas with an 
arbitrary distribution of $r$ by in. However more realistic models 
would require the joint study of routing and MAC, which is beyond 
the scope of this paper.
}

\subsection{Aloha Models --- Time  Added} 
\label{subsec:alohamod} 

We will now consider two time-space scenarios appropriate for slotted
and non-slotted Aloha. 
In both of them the planar locations of MANET
nodes and their receivers $\widetilde\Phi$ {\em remain fixed}.
It is the medium access control (MAC) 
status of these nodes that will evolve differently 
over time depending on which of the following two models is used.

\subsubsection{Slotted Aloha}
\label{sss.slotted}
In this model we assume that the time is discrete, i.e. divided into
slots of length $B$ (the analysis will not depend on the length of
the time-slot) and labeled  by integers $n\in\bbZ$. 
The nodes of $\Phi$ are {\em perfectly synchronized} 
to these (universal) time slots  and send packets according to the
following {\em slotted Aloha policy:
each node, at each time slot independently tosses a coin with 
some bias $p$ which will be referred to as the medium access
probability (MAP); it sends the packet in this time slot if the outcome is
heads and  backs off its transmission otherwise}.
This evolution of the MAC status of each node $X_i$ can be formalized
by introducing its further (multi-dimensional) mark $(e_i(n)
:n\in\bbZ)$,
where  $e_i(n)$  is the medium access indicator of node $i$ at time
$n$; $e_i(n)=1$ if node $i$ is allowed to transmit in the time
slot considered and 0 otherwise.
Following the Aloha principle we assume that $e_i(n)$ are 
i.i.d. (in $n$ and $i$) and independent
of everything else, with $\Pro(e_i(n)=1)=p$.
We treat $p$ as the main parameter to be tuned for slotted Aloha.
We will call the above case {\em the slotted Aloha model}.

\subsubsection{Poisson-renewal Model of Non-slotted Aloha}
\label{sss.renewal}
In this {\em non-slotted Aloha} model all the nodes of $\Phi$
independently, without synchronization, 
send packets of the same duration $B$ 
and then back off for some random time.
This can be integrated in our model by introducing marks
$(T_i(n):n\in\bbZ)$,
where  $T_i(n)$ denotes  the beginning of the $n\,$th transmission
of node $X_i$ with
$T_i(n+1)=T_i(n)+B+E_i(n)$, where 
$E_i(n)$ is the duration of the $n\,$th back-off time of 
node $X_i$. The non-slotted Aloha principle states that 
$E_i(n)$ are i.i.d. (in $i$ and $n$)
independent of everything else. In what follows we assume that  
$E_i(n)$ are  exponential with mean $1/\epsilon$ 
and will consider the  parameter $\epsilon$ as the 
main parameter to be tuned for non-slotted Aloha (given the
packet transmission time $B$). 
More precisely, the lack of
synchronization of the MAC mechanism is reflected in the assumption
that the temporal processes 
$(T_i(n):n\in\bbZ)$ are time-stationary and independent (for different
$i$). Note also that these processes 
are of the {\em renewal} type (i.e., have i.i.d. increments
$T_i(n+1)-T_i(n)$).
For this reason we will call this case the 
{\em Poisson-renewal model for   non-slotted Aloha}.
The MAC state of node $X_i$ at (real) time $t\in\bbR$ can be
described by the on-off process 
$e^{renewal}_i(t)=\ind(T_i(n)\le t<T_i(n)+B\; \text{for some\;} n\in\bbZ)$.

\subsection{Fading and External Noise}
\label{ss.Fading}
We need to complete  our network model by some radio channel
conditions. We will consider the following {\em fading
scenario}: channel  conditions 
vary from one transmission to another and between different
emitter-receiver pairs, but remain fixed for any given transmission.
To include this in our model, we assume a further multidimensional mark
$(\bfF_i(n):n\in\bbZ)$ of node $X_i$ where 
$\bfF_i(n)=(F_i^j(n): j)$ with $F_i^j(n)$ denoting the {\em fading}
in the channel from node $X_i$ to the receiver $y_j$ of node $X_j$ 
during the $n\,$th transmission.
We assume that $F_i^j(n)$ are i.i.d. (in $i,j,n$) and independent 
of everything else.
Let us denote by $F$ the generic random variable of the fading.
We always assume that  $0<\E[F]=1/\mu<\infty$.
In the special case of Rayleigh fading, $F$ is exponential (with
parameter $\mu$).  
(see e.g.~\cite[pp.~50 and~501]{TseVis2005}).
We can  also consider non-exponential cases, which allow
other types of fading to be analyzed, such as e.g. Rician
or Nakagami scenarios or simply the case without any fading (when 
$F\equiv 1/\mu$ is deterministic).

In addition to fading 
we consider a non-negative random variable $W$ independent
of $\widetilde\Phi$ modeling the power of the external  (thermal)
noise. The Laplace transform of $W$ will be denoted by $\calL_W(s)=
\E[e^{-sW}]$. (More generally, we denote by $\calL_U(\xi)$  the Laplace
transform of a random variable $U$.)

The slotted Aloha model described above, when considered in a given
time slot, coincides with the  Poisson Bipolar model with independent
fading considered  in~\cite{BBM06IT}. 
It allows one to derive a simple, explicit evaluation of 
the successful transmission
probability and other characteristics such as
the density of successful transmissions, the mean progress, etc. 

An exact analysis of the Poisson-renewal non-slotted Aloha model, 
albeit feasible, does not lead to similarly closed form expressions. To
improve upon this situation,  
in what follows we propose another  model for the non-slotted case.
It allows the results to be as explicit  as those 
of~\cite{JSAC}, which are moreover 
very close to those of the Poisson-renewal model in a high node density 
regime.

\subsection{Poisson Rain Model for  Non-slotted Aloha}
\label{ss.PoRain}
The main difference with respect to the scenario considered above is
that the nodes $X_i$ and their receivers $y_i$ are not fixed in time.
Rather, we consider a time-space Poisson point process
$\Psi=\{(X_i,T_i)\}$ with $X_i\in\bbR^2$ denoting the location of the
emitter which sends a packet during time interval $[T_i,T_i+B)$
(indexing by $i$ is arbitrary and in particular does not mean
successive transmissions over time).
We may think of node $X_i$ as being ``born'' at time $T_n$ transmitting a packet
during time $B$ and ``disappearing'' immediately after.
Thus the MAC state of the node $X_i$ at (real) time $t\in\bbR$ 
is simply  
$e_i(t)=\ind(T_i\le t<T_i+B)$.

We always  assume that $\Psi$ is homogeneous (in time and space)
P.p.p. with intensity $\lambda_s$. This parameter corresponds to the 
{\em space-time frequency of channel access}; i.e, the number of 
transmission initiations  per unit of space and time.
The points $(X_i,T_i)$ of the space-time P.p.p. $\Psi$ are marked by
the receivers $y_i$ in the same manner as described in
Section~\ref{ss.Bipolar}; i.e, given $\Psi$,  $\{X_i-y_i\}$ are i.i.d
random vectors with $|X_i-y_i|=r$.

{\color{black} This Poisson Rain model is obviously appropriate 
when the nodes are moving very fast, 
thus at each at transmission it is as if the locations of all the nodes are 
re-sampled. Although this model does not generally cover our networks, 
we will show that the Poisson Rain model leads to closed formulas 
and performs similarly to the usual 
model where the locations of nodes are not re-sampled at each transmission. }

Moreover, they are marked by 
 $\bfF_i=(F_i^j:j)$, with $F_i^j$ denoting the fading in the channel
from $X_i$ to the $y_j$ 
(meaningful only if $X_i, X_j$ coexist for a certain time).
We assume that  
  $F_j^j(n)$ are i.i.d. (in $i,j$) and of everything else, with 
the same generic random fading  $F$ as in Section~\ref{ss.Fading}.
We will call the above model the {\em Poisson rain model for
  non-slotted Aloha}.  It can be naturally 
justified by the {\em mobility of nodes}.

\vskip 0.6 cm 

For all the models that we have preented 
we define the channel access probability $\tau$ 
as the probability that a given node transmits a packet a given time.
In the slotted model $\tau=p$, in the non-slotted Poisson renewal model 
$\tau=\frac{B}{B+1/\epsilon}$. In the non-slotted Poisson rain model 
we have $\tau= \tau_s/\lambda$. 

\section{Interference Analysis}
\label{sec:IA}

\resetcounters
\subsection{Path-loss Model}
Let us assume that all transmitters, when authorized by Aloha, emit packets with
unit signal power and that the receiver $y_i$ 
of node $X_i$ receives the power from the node located at $X_j$
(provided this node is transmitting) equal to $F_j^i/l(|X_j-y_i|)$,
where \hbox{$|\cdot|$} denotes
the Euclidean distance on the plane and  $l(\cdot)$
is the path loss function.
An important special case consists in taking 
\begin{equation}\label{simpl.att}
l(u)=(Au)^{\beta} \quad \text{for $A>0$ and $\beta>2$.}
\end{equation} 
%Note that $1/l(u)$ has a pole at $u=0$,
%and thus in particular {\em is not} correct for small distances
%(and hence in particular for small $u$ compared to $1/\sqrt{\lambda}$).
%Another inconvenience of this path-loss model is that
%the total power received at a given location from an 
%infinite Poisson pattern of transmitters has an {\em infinite mean}
%(where averaging is taken over all configurations of transmitters). 
%
%Despite these drawbacks of the path-loss model~(\ref{simpl.att}),
%we will use it as our default model, because it is precise
%enough for sufficiently large values of $u$ and it simplifies many
%calculations.
Other possible choices of path-loss function avoiding the
pole at $u=0$ consist in taking e.g. $\max(1,l(u))$,
$l(u+1)$, or  $l(\max(u,u_0))$.

%\subsection{SINR Condition}
%\label{sec:sinr}
By interference we understand 
the sum of the signal powers 
received  by a given receiver from  all the nodes transmitting in the network
except the receiver's own  transmitter.
\subsection{Interference}
\label{ssec:interference}
\subsubsection{Slotted Aloha }

Let us denote by $I_{i}(n)$ the {\em interference} at receiver $y_i$ at time $n$; 
i.e., {\em the sum of the signal powers 
received by $y_i$ from  all the nodes in 
$\Phi^1(n)=\{X_j\in\Phi: e_j(n)=1\}$ except $X_i$}, namely,
\begin{equation}\label{e.SN-slotted}
I_i(n)=\sum_{X_j\in\Phi^1(n),\, j \ne i }F_j^i(n)/l(|X_j-y_i|)\,.
\end{equation}

\subsubsection{Non-slotted Aloha}
\label{sss.interference}
When transmissions are not synchronized
(as is the case for non-slotted Aloha) the  interference 
%(i.e. the sum of the signal powers 
%received  by a given receiver from  all the nodes transmitting in the network
%except its own emitter) 
{\em may vary during a
given packet transmission} because other transmissions may start or
terminate during it.
In our Poisson-renewal model of Section~\ref{sss.renewal}
this interference process $I_i(n,t)$
during the $n\,$th transmission to node
$y_i$ can be expressed using~(\ref{e.SN-slotted}) with 
$\Phi^1(n)$ replaced by $\Phi_{ren}^1(t)=\{X_j\in\Phi:
e^{ren}_j(t)=1\}$.  
Similarly, in the Poisson rain model of Section~\ref{ss.PoRain},
the interference process, denoted by $I_i(t)$, 
during the (unique) transmission of node $X_i$ conforms to the above
representation~(\ref{e.SN-slotted}) with 
$\Phi^1(n)$ replaced by $\Psi^1(t)=\{X_j\in\Psi:
e_j(t)=1\}$, and $F_j^i(n)$ replaced by $F_j^i$.

Below, we propose two different ways of taking into account the
variation of the  interference during the packet reception.
% in the SINR condition~(\ref{eq:SINR}):
\begin{itemize}
\item Consider  the {\em maximal interference
    value}  during the given transmission
$I_i^{\max}(n)=\max_{t\in [T_i(n),T_i(n)+B]}I_i(n,t)\,$
or $I_i^{\max}=\max_{t\in [T_i,T_i+B]}I_i(t)\,$ for the Poisson-renewal
or the Poisson rain model, respectively. This choice
  corresponds to the situation 
  where bits of information sent within one given packet are not
  repeated/interleaved so that the {interference 
  needs to  be controlled (e.g. through the SINR condition, 
see Section~\ref{sec_sinr}) at any time of the  packet transmission}
(for all  
  symbols) for the reception to be successful.     
\item Taking the  {\em averaged interference value}
over the whole packet duration
$I_i^{mean}(n)=1/B\int_{T_i(n)}^{T_i(n)+B]}I_i(n,t)\,\md t$
or $I_i^{mean}=\int_{T_i}^{T_i+B}I_i(t)\,\md t$ for the Poisson-renewal
or the rain model, respectively,
 corresponds  to a situation where some coding with
  repetition and interleaving of bits on the whole packet duration is used.
\end{itemize}

In what follows we will be able to express, in closed form expressions, 
the Laplace transform of the averaged interference in our non-slotted models.

\subsection{Laplace transform of the interference}

In what follows we consider the interference experienced by an ``extra'' 
receiver added to the network (say at the origin) during the 
reception of a virtual packet which starts at time $0$. 
In slotted Aloha, this will be just the interference $I$ 
received at the origin during slot $0$. For non-slotted Aloha, 
this will be the interference $I^{mean}$ empirically averaged over 
the time interval $[0,B]$.

The general expression of the 
Laplace transform $\calL_I$ of $I$ in the slotted Aloha scheme 
has already been studied.
Here we recall the result assuming a general fading law,
cf~\cite{BBM06IT,blkakl10,blaszczyszyn2013equivalence}. %\cite{lowen1990power}:
\begin{Prop}\label{res.LI}
For the slotted Aloha model with  path-loss function~(\ref{simpl.att})
and a general distribution of fading $F$ with mean~1, we have
\begin{equation}\label{e.LI}
\calL_I(\xi)=\exp\{-\lambda p A^{-2}\xi^{2/\beta}\kappa_{slotted} \E[F^{\frac{2}{\beta}}] \}\,,
\end{equation}
where $\kappa_{slotted}=\pi\Gamma(1-2/\beta)$ is called the {\em spatial contention factor} for 
slotted Aloha\footnote{The term spatial contention factor was 
introduced in~\cite{Haenggi2009}.}. 
In particular 
\squishlist 
\item $\E[F^{\frac{2}{\beta}}]=1 $ in the no-fading scenario $F\equiv1$,
\item $ \E[F^{\frac{2}{\beta}}]=2 \Gamma(2/\beta)/\beta$ with Rayleigh
  fading ($F$ exponential) and $\mu=1$. 
\item $\E[F^{\frac{2}{\beta}}] = \exp(\sigma^2(2-\beta)/\beta^2)$ with $F$ log-normal shadow-fading of mean 1\footnote{$F=\exp(-\sigma^2/2+\sigma Z)$ with $Z$ standard 
normal random variable.}.
\item $\E[F^{\frac{2}{\beta}}] =  \frac{\Gamma(k+2/\beta)}{\Gamma(k)2^{2/\beta}}$ with $F$ a Nakagami distribution\footnote{the density of $F$ is $f(x,k)= \frac{k^k}{\Gamma(k)} x^{k-1}e^{-kx}$, note that this corresponds to Nakagami $(k,1)$ fading.}. 
\squishend 
\end{Prop}
\medskip
%%The constant $\kappa$  was evaluated in~\cite{BBM03_allerton}
%%for  Rayleigh fading and  in~\cite{Haenggi2009}, for the  no-fading
%%scenario, where the name {\em spatial contention factor}
%%was proposed for this constant. 

Regarding the distribution of the averaged interference $I^{mean}$ in
non-slotted Aloha, we have the following {\em new general} result. 

\begin{Prop}\label{fact.nonslotted_kappa}
Let us consider the Poisson rain model  of non-slotted Aloha with space-time
intensity of packet transmissions $\lambda_s=\lambda\tau$
and the path-loss function~(\ref{simpl.att}). Let us assume  a general distribution of
fading $F$. 
Then the Laplace transform $\calL_{I^{mean}}(\xi)$ of the averaged
interference $I^{mean}$ is given by~(\ref{e.LI}) 
with  the  spatial contention factor $\kappa=\kappa_{non-slotted}$
equal to:
$$\kappa_{non-slotted}=\frac{2\beta}{2+\beta}\kappa_{slotted}\,,$$ 
where $\kappa_{slotted}$ is the spatial contention factor 
evaluated for slotted Aloha under the same channel assumptions.
\end{Prop}

\begin{proof} 

By~(\ref{e.SN-slotted}) and by the definition of $I^{mean}$ (regarding 
the interference experienced by an extra receiver added to the
network at the origin during the reception of a virtual 
packet which starts at time 0), if we 
exchange the order of integration
and summation we can express $I^{mean} $ in the following form:
$$I^{mean}=
\sum_{X_j\in\Phi}F_j^0H_j/l(|X_j|)\,,$$
where $H_j=\frac1B\int_{0}^{B}\ind(X_j\text{\;emits at
  time\;}t)\,dt$. In the Poisson rain model we have
$\ind(X_j\text{\;emits at  time\;}t)=\ind(t-B\le T_j\le t)$, 
where $T_j$ is the time at which $X_j$ starts emitting. Integrating
the previous function we obtain $H_j=h(T_j)$, where $h(s)=(B-|s|)^+/B$ and 
$t^+=\max(0,t)$. Consequently, for the Poisson rain model represented by
Poisson p.p. $\Psi=\{X_i,T_i\}$
(cf. Section~\ref{ss.PoRain})
the averaged interference at the typical transmission receiver is
equal in distribution to:
$$I^{mean}\;\buildrel{\text distr.}\over=
\sum_{X_j,T_j\in\Psi}F_{j}^0 h(T_j)/l(|X_j|)\,,$$
where $F_j^0$ are i.i.d. copies of the fading.
Using the general expression for the Laplace transform of the
Poisson shot-noise (see e.g.~\cite[Prop.~2.2.4]{FnT1}), 
for the path-loss function~(\ref{simpl.att}) we obtain: 
\begin{align*}
&\calL_{I^{mean}}(\xi)\\
&=\exp\Bigl\{-2\pi\lambda_s\int_{-\infty}^\infty\!\!\!\int_0^\infty
\!\!\!r\Bigl(1-\calL_F\bigl(\xi h(t)(Ar)^{-\beta}\bigr)\Bigr)
\,dr\,dt\Bigr\}\,,
\end{align*}
where $\calL_F$ is the Laplace transform of~$F$. 
Substituting $r:= Ar(\xi h(t))^{-1/\beta}$ for a given fixed $t$ 
in the inner integral we
factorize the two integrals and obtain
$\calL_{I^{mean}}(\xi)=\exp\{\lambda_s A^{-2}\xi^{2/\beta}\zeta\eta\}$,
where $\zeta=\int_{-\infty}^\infty (h(t))^{2/\beta}\,dt$
and $\eta=2 \pi\int_0^\infty
r(1-\calL_F(r^{-\beta}))\,dr=\kappa_{slotted}
\E[F^{\frac{2}{\beta}}]
$.
A direct calculation yields $\zeta=2\beta/(2+\beta)$.
This completes the proof.
\end{proof} 

\begin{figure}[h]
\centering\includegraphics[width=0.9\linewidth]{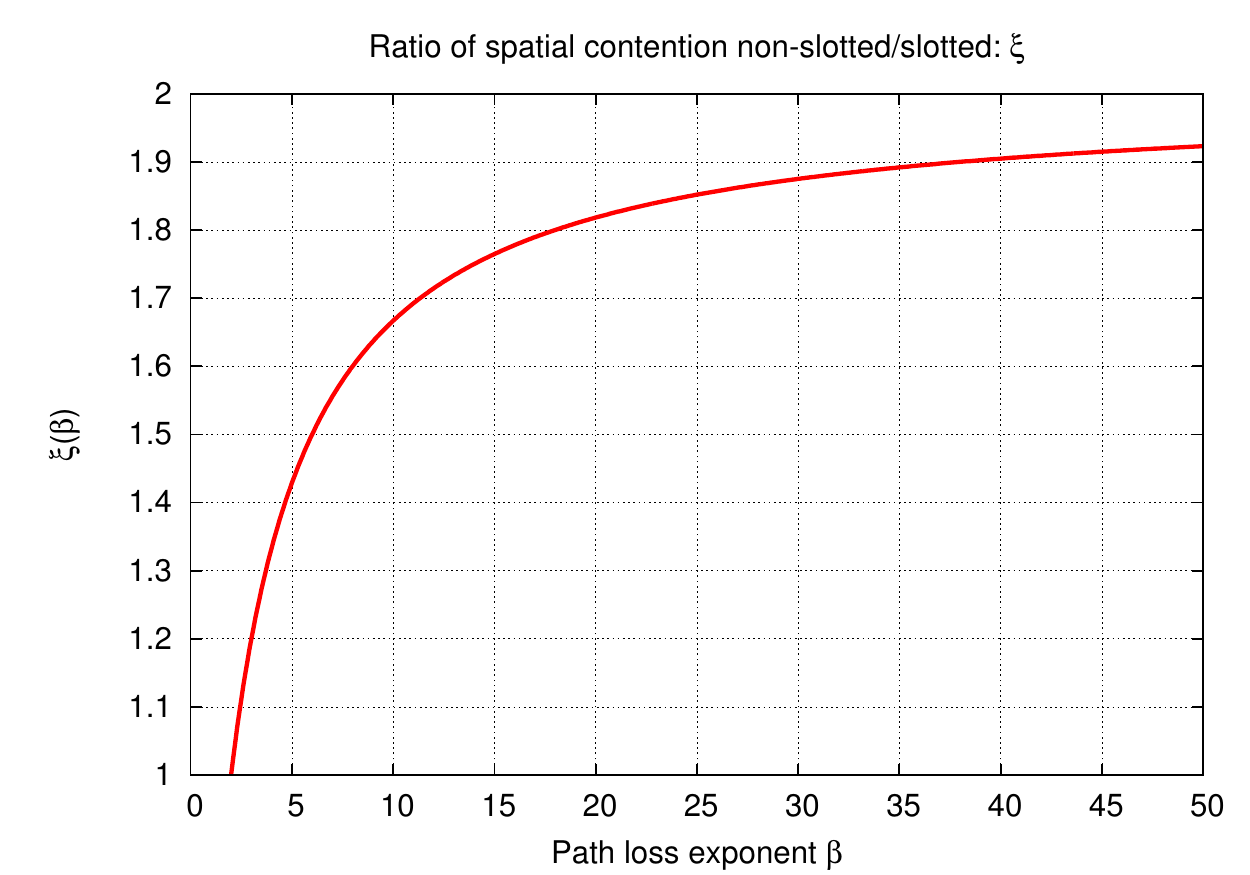}
\caption{Ratio of the spatial contention parameters between non-slotted and
slotted Aloha models versus $\beta$.}
\label{fig.zeta}
\end{figure}

\begin{rem}\label{r.price_nonsyncro}
Recall that 
$$\zeta=\zeta(\beta)=\frac{\kappa_{non-slotted}}{\kappa_{slotted}}=2\beta/(2+\beta)
$$
is the ratio of the spatial contention parameters between non-slotted and
slotted Aloha models. 
It  can be seen as the {\em contention cost of non-synchronization in
  Aloha} (cf Remark~\ref{r.optimal_Aloha} below).
We plot its value as a function of $\beta$ in Figure~\ref{fig.zeta}.
Note that in the free-space propagation model (where  $\beta=2$)
it is equal to~1 (which means that the interference distribution
%and therefore the coverage  probability, 
in slotted and non-slotted Aloha are the same. 
Moreover, $\zeta(\beta)$ increases with the
path-loss exponent and asymptotically (for $\beta=\infty$) approaches
the value~2 (as conjectured in~\cite{BBPMnsaloha09}).  
\end{rem}

\subsection{Interference in the Poisson-Renewal Model}
%\toself TODO in function of the remaining space. If not enough space, 
%can given only the main ideal and mode details in an extended version 
%deposed in some open archive.
We now consider the mean interference $I^{mean}=I^{mean}_0(0)$ 
experienced by the receiver located at the origin~0, during the
transmission of the packet, which starts at time $T_0=0$.
%under the $\Pro^{0,T_0(0)=0}$. 

\begin{Prop} \label{prop_li_g}
The Laplace transform of the interference $I^{mean}$ in the
Poisson-renewal model with arbitrary distribution of fading  $F$
satisfying  $\E[F^{2/\beta}]<\infty$ is given by the following expression: 
\begin{align*}
&\calL_I^{mean}(\xi) = \exp \Big[-\int_0^{\infty}\Big(1-
\frac{\epsilon B (e^{-\xi/v} -e^{-B \epsilon }) }{(1+\epsilon B)(\epsilon B- \xi /v)}\\[1.5ex]
&-\frac{(\epsilon B)^2}{(1+\epsilon B)(\xi /v- \epsilon B)} \int_0^1 
\!\!\!\!\!e^{\epsilon Bt} (1-e^{(\xi /v-\epsilon B)(t-1)})   \tilde{\calL}(\xi t/v) \md t \\[1.5ex]
&- \frac{\epsilon B e^{-\epsilon B}}{(1+\epsilon B)} \int_0^1 e^{\epsilon Bt} \tilde{\calL}(\xi t/v) \md t - \frac{e^{-\epsilon B}}{(1+\epsilon B)}   \Big) \Lambda(dv) \Big]\,,
\end{align*}
where 
\begin{align*}
\Lambda(dv)&=\frac{2 \lambda \pi \E[F^{2/\beta}]}{A^2 \beta}
v^{\beta/2-1}\,\md v \\
\tilde{\calL}(\eta) &=\frac{1}{\E[F^{2/\beta}]} \E[F'^{2/\beta}\calL_F(\eta / F')]  \\
\end{align*}

\noindent  $F$ and $F'$ being independent with the law of the fading. 

%%\big(1-\frac{(\epsilon B)^2}{(1+\epsilon B)(\xi /v- \epsilon B) \int_0^1 e^{\epsilon Bt} \tilde{\calL} \big) \delta(dv) \bbig]
\end{Prop}
The {\em proof} of this proposition in given in the Appendix. %%~\ref{appendix}. 
%This formula is interesting 
%because it is very general the only assumption of the fading being
%that $\E[S^{2/\beta}]<\infty$. 
The following two results consider two
special cases: constant ($F \equiv 1$) and Rayleigh ($F$ exponential)
fading.

\begin{Prop} \label{prop_li_f1}
Under the assumptions of Proposition~\ref{prop_li_g}, with $F\equiv1$
$$ \calL_{I^{mean}}(\xi )= \exp\Bigl\{- \frac{2 \pi \lambda}{A^2 \beta} \int_0^{\infty} i(\xi/v) v^{2/\beta-1} \md v \Bigr\}$$
with 
\begin{align*}
&i(\xi/v)= 1-\frac{ e^{-\epsilon B } (  \epsilon B e^{\epsilon B -\xi/v}  - \xi/v  ) }{(1+\epsilon B)(\epsilon B - \xi/v) }\\[1.5ex]
&-\frac{\epsilon B(\epsilon^2 B^2 e^{-\xi/v} - \epsilon B
  e^{-\xi/v}\xi/v - e^{-\xi/v}\xi/v + e^{-\epsilon B } \xi/v
  )}{(1+\epsilon B)(\epsilon B - \xi/v)^2}  \,.
\end{align*}
\end{Prop}
\begin{proof}
When $F\equiv 1$ we have $\tilde{\calL}(\eta)= e^{-\eta}$
and $\Lambda(dv)=\frac{2 \lambda \pi}{A^2 \beta} v^{\beta/2-1}\md v$.
 %Then $F_2$ and $F_1$ can exactly 
%computed with a closed formula. This provides the value of
%$i(\xi/v)$.  
Inserting the values of $\tilde{\calL}(\eta)$ and $\Lambda(dv)$ 
in the expression given in  Proposition~\ref{prop_li_g} and
after some algebra, one obtains the result. 
\end{proof}
\begin{Prop} \label{prop_li_expo}
Under the assumptions of Proposition~\ref{prop_li_g}, with exponential
fading $F$ of mean $1$ (which corresponds to Rayleigh fading) we have:
%% \begin{eqnarray*}
%% \lefteqn{ \calL_{I^{mean}}(\mu Tl(r) ) = \exp\Bigl\{-\lambda \int_{x\in\mathbb{R}^2}\Bigl(1-\frac{\epsilon B}{1+ \epsilon B} } \\
%% &&
%% \times \frac{1}{B} \int_0^B \frac{1}{ 1+\frac{(B-t)Tl(r)}{Bl(|x|)}} \int_0^{\infty} \frac{\epsilon e^{-\epsilon s}}{ 1+\frac{(t-s)^+Tl(r)}{Bl(|x|)}} \md s \md t \\
%% && 
%% - \frac{1}{1+ \epsilon B} \int_0^{\infty} \frac{\epsilon e^{-\epsilon s}}{ 1+\frac{(B-s)^+Tl(r)}{Bl(|x|)}} \md s    \Bigr)\md x  \Bigr\}\,.     
%% \end{eqnarray*}
\begin{align*}
&\calL_{I^{mean}}(\xi )=\exp\Bigl\{-\lambda
\int_{x\in\mathbb{R}^2}\Bigl(1
- \frac{1}{1+ \epsilon B} \int_0^{\infty} \frac{\epsilon e^{-\epsilon s}}{ 1+\frac{(B-s)^+\xi}{Bl(|x|)}} \md s\\[1.5ex]
&-\frac{\epsilon B}{1+ \epsilon B} 
\frac{1}{B} \int_0^B \frac{1}{ 1+\frac{(B-t)\xi}{Bl(|x|)}} \int_0^{\infty} \frac{\epsilon e^{-\epsilon s}}{ 1+\frac{(t-s)^+\xi}{Bl(|x|)}} \md s \md t \Bigr)\md x  \Bigr\}\,.     
\end{align*}

\end{Prop} 

The proof of this formula can be found in \cite{BBPMnsaloha09}.  

%\vspace{2. cm}

\section{SINR coverage Analysis}
\label{sec_sinr}
In the slotted Aloha model 
it is natural to assume  that transmitter $X_i$ 
{\em covers} its receiver $y_i$ in time slot $n$ if
\begin{equation}\label{eq:SINR}
\mbox{SINR}_i(n)=\frac{F_i^i(n)/l(|X_i-y_i|)}{W+I_i(n)}\ge T\,,
\end{equation}
where  $T$ is some SINR threshold. 
When  condition~(\ref{eq:SINR}) is satisfied we say that $X_i$ can be {\em
  successfully  received} by $y_i$ 
or, equivalently,  that $y_i$ {\em is not in  outage} with respect to
$X_i$ in  time slot $n$.
We will say that in {\em non-slotted Aloha with maximal interference
  constraint}, 
$X_i$ can be successfully  received by $y_i$
(in time slot $n$ in the case of  the Poisson-renewal model), 
if condition~(\ref{eq:SINR}) holds with 
$I_i(n)$ replaced by  $I_i^{\max}(n)$ or $I_i^{\max}$  in the
Poisson-renewal or the Poisson rain 
models, respectively.
Similarly,  we will say that in {\em non-slotted Aloha with average
 interference  constraint},
$X_i$ can be successfully  received by $y_i$
(in time slot $n$ in the case of  the Poisson-renewal model), 
if condition~(\ref{eq:SINR}) holds with 
$I_i(n)$ replaced by  $I_i^{mean}(n)$ or $I_i^{mean}$  in the Poisson-renewal
or Poisson rain models, respectively.

We will see that the coverage 
probability with the average
 interference  constraint (in 
both the Poisson-renewal and the Poisson
rain model), can be easily derived from the Laplace transform of interference 
$I^{mean}$. We have not been able to derive closed formulas when 
the maximal interference constraint is used; this case is studied by simulations
in Section~\ref{sec:comp}.

\subsection{Coverage probability}
Let us define the coverage probability $p_c$ in the three
models considered as the probability that the SINR value in the 
typical transmission exceeds the threshold $T$. More formally,  
we have
\begin{align*}
p_c&:=\Pro^{(0,0)}\{\,\text{SINR}_0(0)\ge T\,\}\\
&=\Pro^{(0,0)}\{\,F_0^0(0)\ge l(r)T(W+I_0(0)\,\}
\end{align*}
where $\Pro^{(0,0)}$ denotes the Palm probability of $\Phi$ (given 
a node $X_0=0$ at the origin) and given that it starts its transmission
at time 0 (indexed by $n=0$), and where $I_0(0)=I_0^{mean}(0)$ for the
non-slotted case.

\begin{Res}
For the slotted and Poisson-rain non-slotted  Aloha models with the path-loss function~(\ref{simpl.att})
and Rayleigh fading of mean  $1/\mu=1$, we have 
\begin{equation}\label{eq:p_r_lambda_1}
p_c=\exp\Bigl\{-\lambda\tau r^2T^{2/\beta}\kappa\Bigr\}\,,
\end{equation}
\label{e.p_c_non-slotted}
where 
\squishlist 
\item $\kappa=\kappa_{slotted}\,\Gamma(2/\beta+1)=\pi\Gamma(1-2/\beta)\Gamma(1+2/\beta)/\beta =
2\pi\Gamma(2/\beta)\Gamma(1-2/\beta)/\beta$ and $\tau
  =p$ for slotted Aloha,  
\item $\kappa=\kappa_{non-slotted}\, \Gamma(2/\beta+1)=4\pi\Gamma(2/\beta)\Gamma(1-2/\beta)/(2+\beta)$ for the non-slotted Poisson Aloha model.
\squishend 
\end{Res}
\medskip

\begin{proof}
The basic observation allowing explicit analysis of the
coverage probability for all our  Poisson models of 
Aloha is that the  distribution of the interference  $I_0(0)$
and $I^{mean}_0(0)$
and  under $\Pro^{(0,0)}$ is independent of $W$ and $F_0^0(0)$, and 
corresponds 
to the  distribution of the interference  ``experienced'' by  an extra
receiver in the respective model, studied in Section~\ref{sec:IA}.
This is a consequence of Slivnyak's theorem; cf.~\cite[Theorem 1.4.5]{FnT1}.
Consequently, for all three  (slotted, non-slotted rain or renewal)
Aloha models, in the special case of Rayleigh fading we have 
$p_c=\calL_W(Tl(r))\calL_{I}(Tl(r))$, where $I$ is the 
interference in the respective model ($I=I^{mean}$ for the non-slotted
models).  By Propositions~\ref{res.LI} and~\ref{fact.nonslotted_kappa} we have the above result. 
\end{proof}

For a general distribution of fading the evaluation of $p_c$
from the Laplace transform $\calL_{I}$  (or  $\calL_{I^{mean}}$) 
is not so straightforward.
Some integral formula, based on the Plancherel-Parseval theorem,
was proposed in~\cite{JSAC} when $F$ has a square
integrable density. This approach however does not apply
to the no-fading case $F\equiv1$. Here we suggest another, numerical  approach,
based on the Bromwich contour inversion integral and developed
in~\cite{AbateWhitt95}, which is particularly efficient in this case.
\begin{Prop}\label{fact.Bromwich}
For  the (slotted, non-slotted rain and renewal) Aloha model with constant fading $F\equiv1$ we have
\begin{equation}\label{eq:Bromwich}
p_c=\frac{2\exp\{\delta/(Tl(r))\}}{\pi}
\int_{0}^\infty \calR\Bigl(\frac{1-\calL_{I}
(\delta+iu)}{\delta+iu}\Bigr)\cos(uT)\,du\,,
\end{equation}
where $\delta>0$ is an arbitrary constant
and 
$\calR(z)$ denotes the real part of the complex number $z$, with $\calL_I$ being the Laplace 
transform of the (mean) interference in the respective Aloha model.
\end{Prop}
As stated in~\cite{AbateWhitt95}, the integral in~(\ref{eq:Bromwich}) 
can be numerically evaluated using the trapezoidal rule and the Euler
summation rule can be used to truncate the infinite series;
the authors also explain how to set $f\delta$ in order to control the
approximation error.

\subsection{Space-time efficiency}
\label{subsec:spacetime}

Spatial density of successful transmission is usually defined as  
the average number of successful transmissions per unit of time and 
space. This performance characteristic, 
introduced in~\cite{BBM06IT} and 
widely accepted since then, can be considered as a measure of spatial 
throughput of the network. We will use this term in this paper. 
By the standard Campbell average formula, the  spatial 
throughput is $d =\lambda \tau p_c$. 

In what follows we optimize and compare 
this quantity for slotted and non-slotted 
Poisson-rain Aloha models assuming exponential Rayleigh fading.  

The following result follows immediately from~(\ref{e.p_c_non-slotted}).
\begin{Prop}\label{r.opt_d_succ}
We assume that there is no noise $W=0$,  Rayleigh fading and
path-loss~(\ref{simpl.att}). The  maximum value of the  
spatial throughput is $d_{max} =1/(e r^2T^{2/\beta} \kappa)$ in 
slotted and non-slotted rain Aloha. This maximum 
is attained for the space-time density of channel access
$\lambda\tau=1/ r^2T^{2/\beta} \kappa $.
Moreover, given the spatial density of nodes
$\lambda$ the optimal mean channel-access-time-fraction $\tau$ per
node  $\tau_{\max}$ for  $d_{suc}$
is equal to 
$$\tau_{\max}=\frac1{\lambda r^2T^{2/\beta} \kappa}\,.$$
if $\lambda>1/r^2T^{2/\beta} \kappa$
and $1$ (interpreted as no back-off; i.e., immediate retransmission)
otherwise.
\end{Prop} 
\medskip

\begin{rem}
Since $\kappa$ is larger for non-slotted Poisson rain Aloha than 
for slotted Aloha when we optimally tune the two protocols we have 
$\tau_{\max}^{slotted} > \tau_{\max}^{non-slotted}$. This means
that optimally tuned  non-slotted  Aloha occupies less channel
than optimally tuned slotted Aloha. 
\end{rem}

The following result compares the spatial throughput
for slotted and non-slotted Aloha, when both are optimized
\begin{Prop}\label{r.optimal_Aloha}
Under the assumptions of Result~\ref{r.opt_d_succ},
the ratio of the spatial throughput for 
slotted and non-slotted Poisson rain Aloha is:

$$\frac{d_{max}^{slotted}}{d_{max}^{non-slotted}   }=
\frac{\kappa_{non-slotted}}{\kappa_{slotted} }=\frac1{\zeta(\beta)}
=\frac{2 \beta}{\beta+2}\,.$$     
\end{Prop}
In Figure~\ref{fig.cpdebitop_bvar1} we present this good-put ratio
for the optimized systems as a function of $\beta$
(note that it does not depend on other parameters such as $\lambda,r,T$).   
We observe that for small values of path-loss exponent $\beta$ (close
to $2$) the performances of slotted and non-slotted Aloha are similar, but for
large values of $\beta$ non-slotted Aloha performs significantly worse 
than slotted Aloha.
However even for $\beta=6$ the ratio still remains significantly
larger than $50\%$  
foreseen by the  'well-known' (simplified collision) model (See \cite[Section~4.2]{Bertsekas01}).

\subsection{Energy efficiency}
\label{subsect:enereff}

Note that both slotted Aloha and non-slotted Aloha (in the rain model), 
when optimally tuned
for the sptatial throughput $\tau = \tau_{max}$, 
have the same probability of successful transmission $p_c=1/e$. 
This means that  both schemes
exhibit the same  energy efficiency. If one assumes 
that each  transmission requires one energy unit, 
the number of successful transmissions per unit of 
energy consumed is $p_c=1/e$. Let us remark that this comparison does 
not take into
account the energy used to maintain synchronization 
in the slotted scheme. Taking this fact into account 
we can say that non-slotted Aloha out-performs slotted Aloha in 
terms of energy efficiency.

\section{Numerical results} 
\label{sec:comp}

In this section we validate our models by comparing
them with simulation results. {\color{black} 
The simulations are performed with random locations 
of the nodes drawn 
at the beginning of the simulation but not re-drawn
at each transmission. Thus, in principle, 
the simulations are more aligned with the Poisson renewal 
model.} 
We also compare 
the performance of slotted and non-slotted Aloha.   
If not otherwise defined we use the following numerical assumptions :
$\lambda = 1$, $r=1$, $T=10$ and $\beta=4$. 
The simulations are carried out in a square of 300 m $\times$ 300 m with 
the same numerical assumptions. 
{\color{black} Let us recall that $\lambda$ 
is the node density, $r=1$ is the mean distance between the sender and 
receiver node, $T=10$ is the capture threshold and $\beta=4$ is the path-loss 
exponent. We use  $r=1$ which is, in this network, twice the  mean distance 
from the current node to its closest 
neighbor, $T=10$ is a default assumption when no coding (or other advanced 
technique) is  used. The assumption $\beta=4$ corresponds to the 
path-loss observed in dense urban scenarios. }

\subsection{Non-slotted Aloha,  Renewal versus Rain model} 

\begin{figure}[!t]
\centering\includegraphics[width=0.9\linewidth]{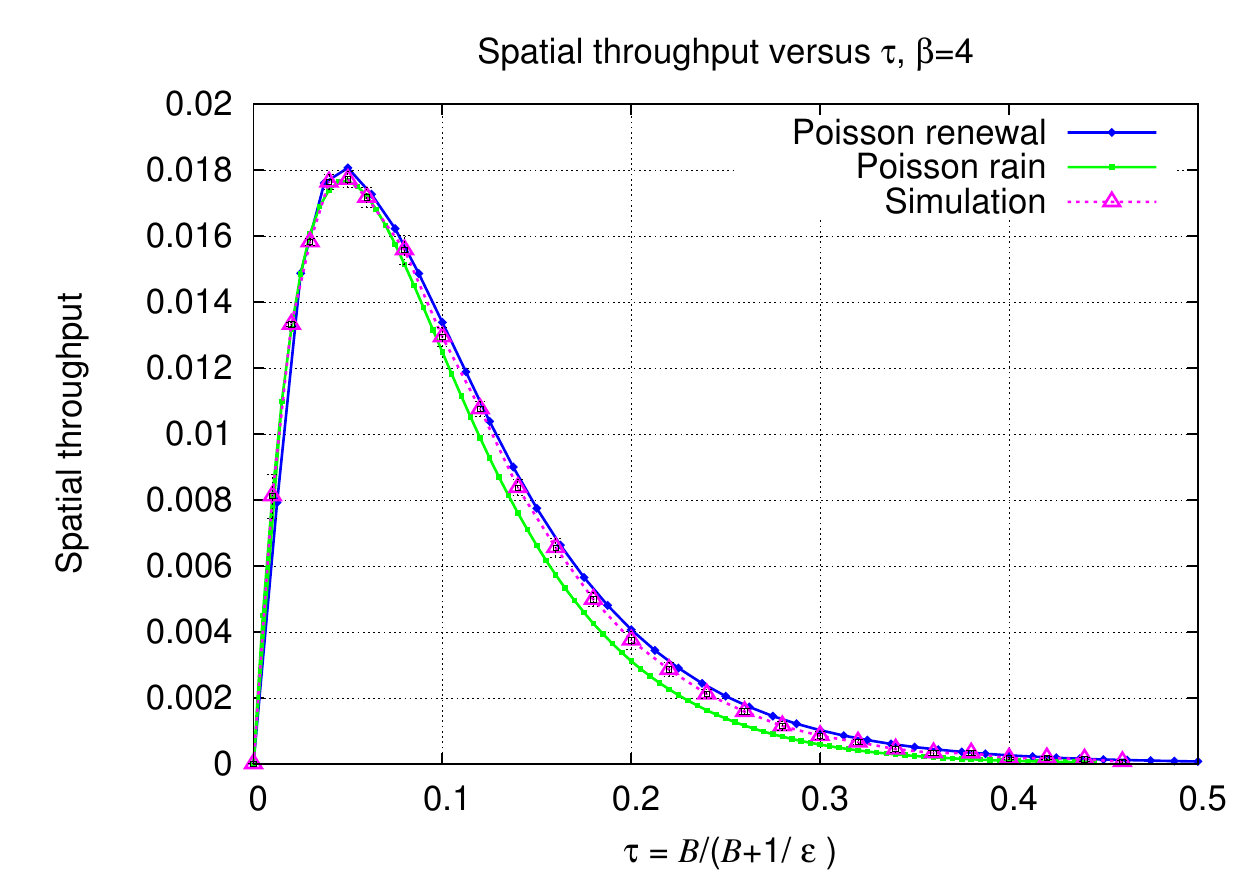}
\caption{Spatial throughput versus $\tau=\frac{B}{B+1/\epsilon}$. Comparison of the Poisson-renewal and the Poisson rain models with simulation results. Rayleigh fading. }
\label{fig.ns_aloha_modsim1}
\end{figure}

In this section we validate our Poisson rain and renewal models 
by comparing
the spatial throughput  $\lambda \tau p_c$ obtained 
with these two models and with simulations. 
We adopt Rayleigh fading. 
For the rain model we have a closed formula 
but for the renewal model $ p_c$  is evaluated numerically using Proposition~\ref{fact.Bromwich} 
and the Laplace transform derived from Proposition~\ref{prop_li_g}. 
We use the same numerical assumptions as previously:
$\lambda = 1$, $r=1$,  $T=10$ and $\beta=4$. 
In Figure~\ref{fig.ns_aloha_modsim1} we observe a very good
matching of the two models with simulations. {\color{black} 
As already stated we are interested in scenarios with low mobility. 
Thus, in the simulations, the nodes' locations are not re-sampled 
at each transmission and the simulations are closer to 
the Poisson renewal model than to the Poisson rain model.  
Simulations with high mobility closer to the Poisson rain model are not 
presented in Figure~\ref{fig.ns_aloha_modsim1}) as they would have 
been aligned given with the Poisson Rain model.}

\medskip
We perform the same comparison between the two models and the simulations 
for $\beta=5$ and $\beta=3$. For $\beta=5$ the matching of the 
two models and the simulation is perfect. For $\beta=3$ the two 
models provide the same results whereas the simulations give a larger 
spatial of throughput. This can be explained by the fact that the simulation 
network is of finite length, and the border effects have a stronger
impact  for small values of $\beta$.  

\begin{figure}[!t]
\centering\includegraphics[width=0.9\linewidth]{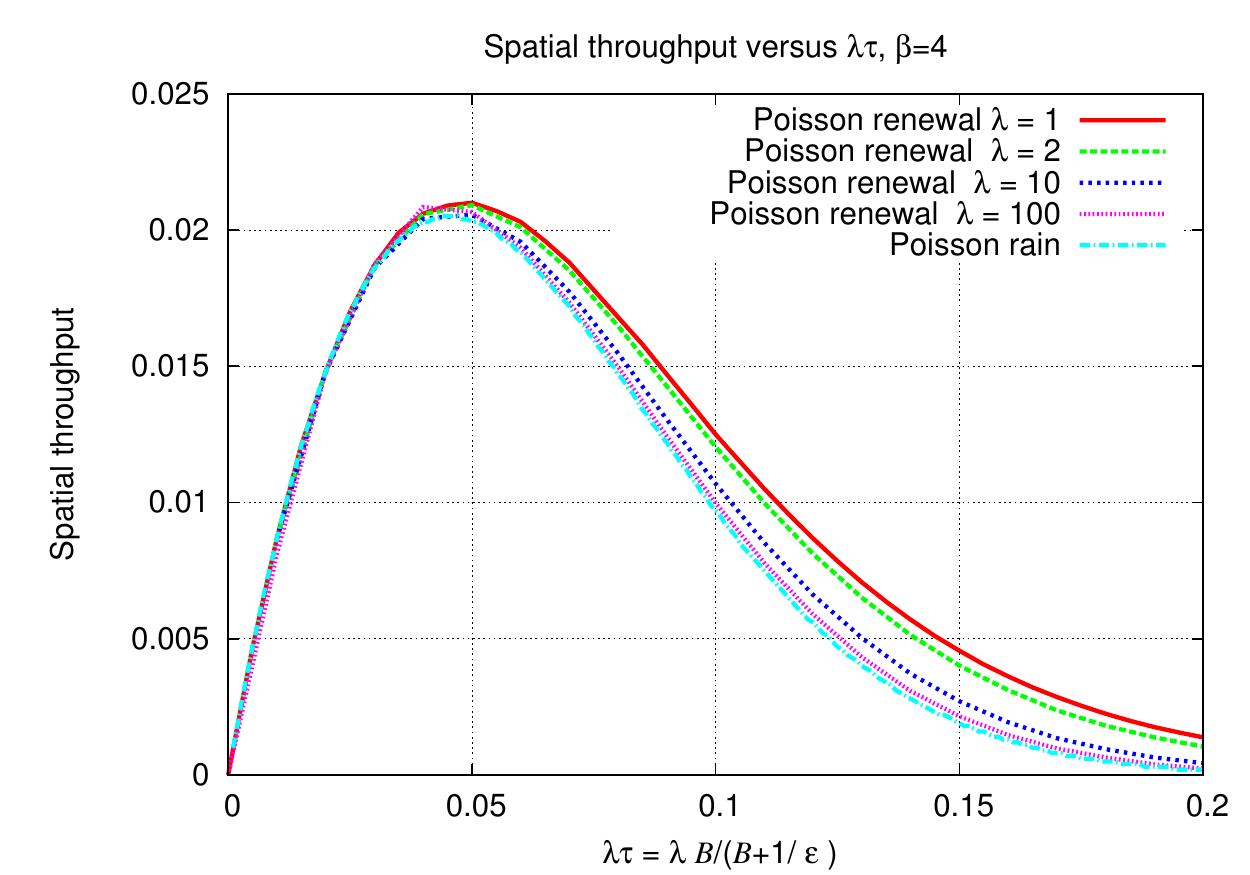}
\caption{Spatial throughput versus $\tau=\frac{B}{B+1/\epsilon}$. Convergence of the Poisson-renewal model towards the Poisson rain model.}
\label{fig.ns_aloha_f2}
\end{figure}

\medskip
We then compare the Poisson rain and renewal models with no fading i.e. 
$F\equiv 1$. The result of this comparison is shown 
in Figure~\ref{fig.ns_aloha_f2}. We vary $\lambda$ from 1 to 100 
in the renewal model. We observe that the rain and the renewal models 
for $\lambda = 100$ show similar performances. This is 
because when $\lambda$ is large, the transmitting node changes 
at every transmission, which is very close to the Poisson rain model.  
For small values of $\lambda$ the two models show slightly different 
results. 
\medskip

\subsection{Slotted versus non slotted Aloha}

\begin{figure}[!t]
\centering\includegraphics[width=0.9\linewidth]{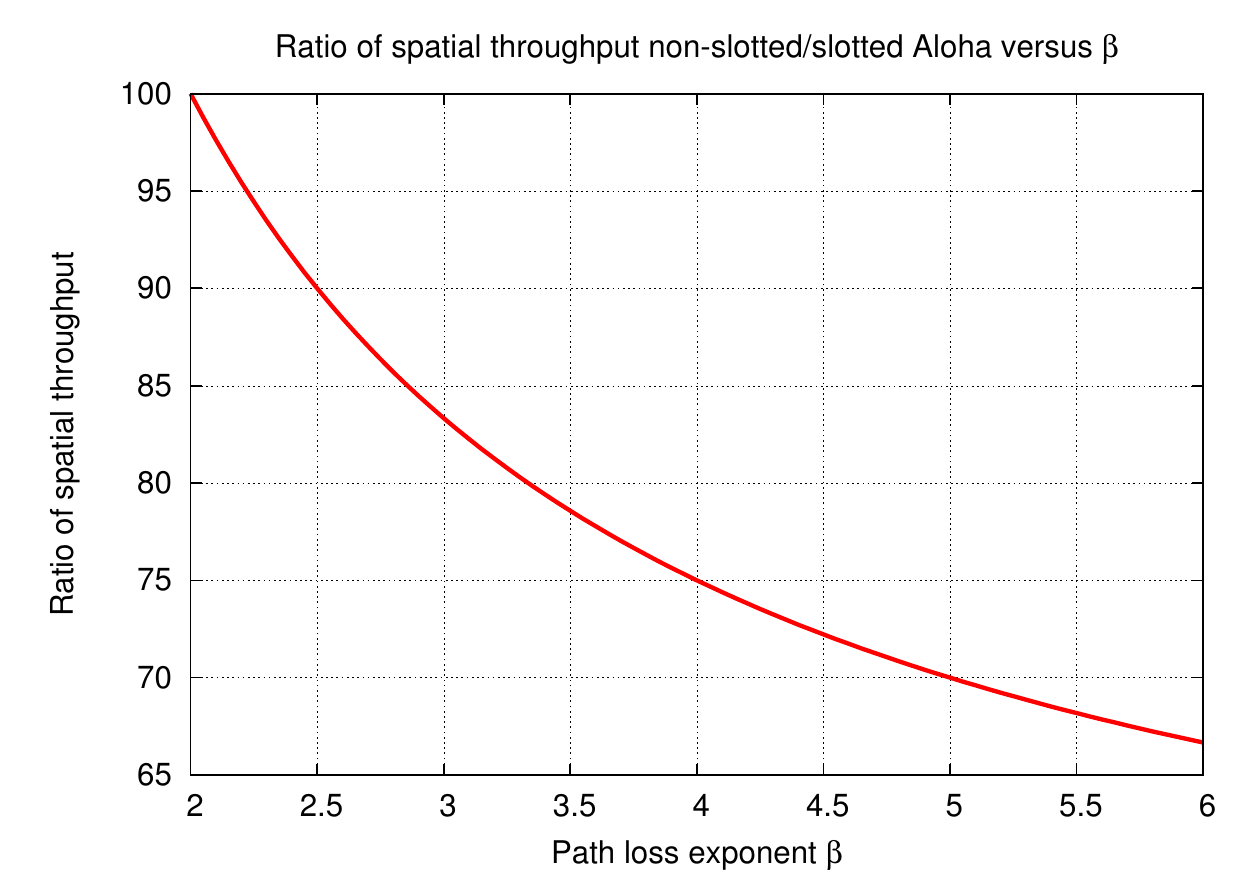}
\caption{The ratio (in \%) of the good-put offered by non-slotted
  Aloha with respect to slotted Aloha, when both are optimized so as
  to  to maximize the spatial throughput as a function 
of the path loss exponent $\beta$. It is  equal to the inverse of
the contention cost $1/\zeta(\beta)$ and does not 
depend on any other parameter.} 
\label{fig.cpdebitop_bvar1}
\end{figure}

In Figure~\ref{fig.cpdebitop_bvar1}, we present the ratio 
of spatial throughput
transmissions for non-slotted Aloha and slotted Aloha. 
We use the analytical model of slotted-Aloha and the Poisson rain 
model for non-slotted Aloha. 
We optimize 
non-slotted Aloha and slotted Aloha with respect to $p$ and $\tau$. 
We observe that for small values of the path loss exponent (close to 2) 
the performances of slotted and non-slotted Aloha are similar,
but for large values of $\beta$ slotted Aloha significantly 
outperforms non-slotted Aloha.

\begin{figure}[!t]
\centering\includegraphics[width=0.9\linewidth]{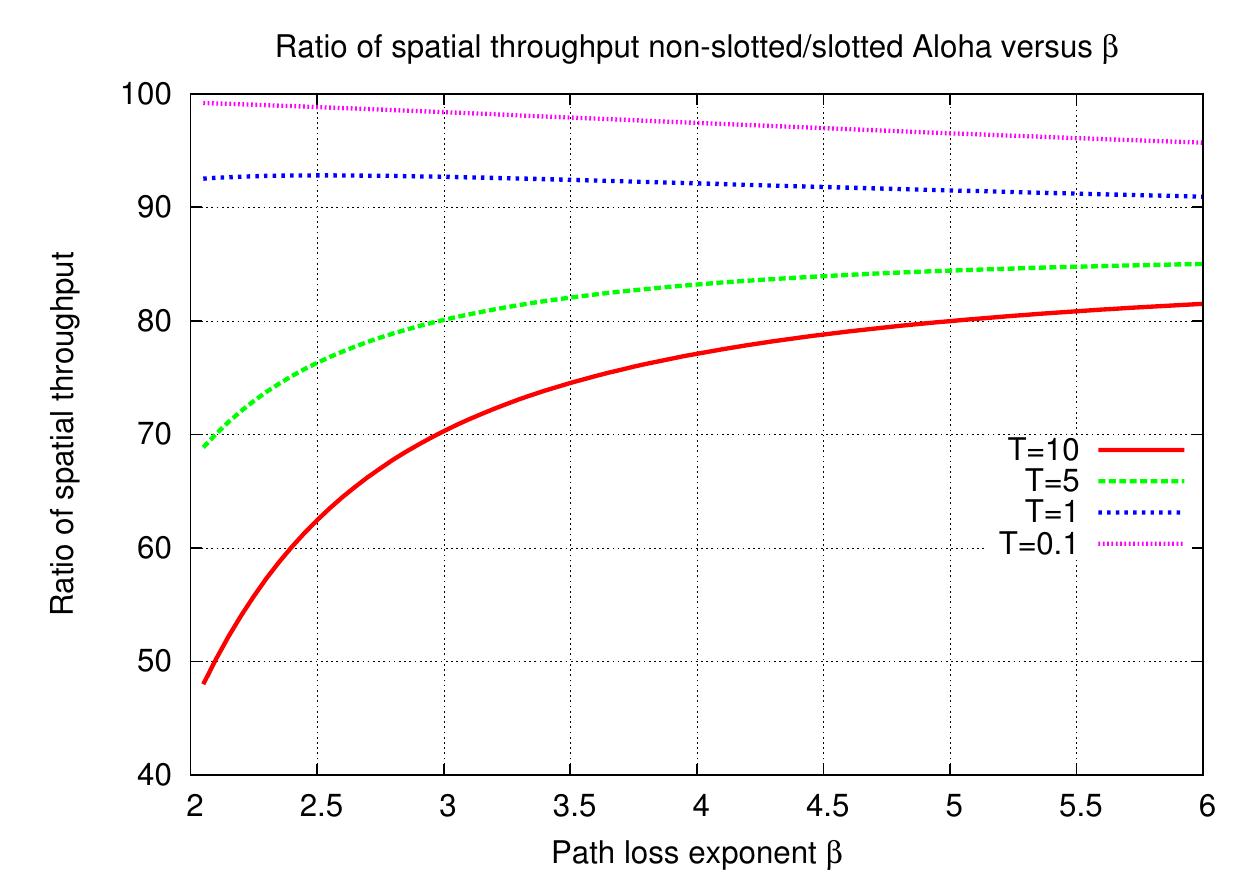}
\caption{The ratio (in \%) of the good-put offered by non-slotted
Aloha with respect to slotted Aloha, as a function of the path loss
  exponent $\beta$, for various choices of the SINR threshold $T$;
other parameters are $p=\frac{\epsilon}{1+\epsilon}=0.05$, $\lambda = 1$, $r=1$.}
\label{fig.cpdebit_bvar1}
\end{figure}

In Figure~\ref{fig.cpdebit_bvar1} we compare the spatial throughput 
for slotted Aloha and non-slotted Aloha when 
$\tau=p=\frac{\epsilon}{1+\epsilon}=0.05$. We observe 
that for small values of the path loss exponent (close to 2) 
the performances of slotted and non-slotted Aloha are similar 
when $T$ is small but that slotted Aloha significantly outperforms 
non-slotted Aloha for medium to large values of $T$. 
For large values of $T$, slotted Aloha still outperforms non-slotted 
Aloha but the difference remains less than when $T$ is small, 
non-slotted Aloha provides more than 80\% of slotted Aloha throughput. 
We observe that in contrast to the case where the protocols 
are tuned to offer the maximum throughput, the ratio of throughput 
for non-slotted/slotted Aloha does depend on $T$ (and of course 
on $\tau$ or $p$).

%% \begin{figure}[!h]
%% \centering\includegraphics[width=0.90\linewidth,height=0.65\linewidth]{figures/ns_aloha_mean_bvar.pdf}
%% \caption{Density of successful transmissions versus path loss exponent
%%  $\beta$. Slotted Aloha and non-slotted Aloha (mean and maximal interference constraint) are tuned to maximize the density of successful transmissions. }
%% \label{fig.ns_aloha_mean_bvar}
%% \end{figure}

\subsection{Mean Versus Maximum Interference Constraint in SINR}
\label{ss.mean-max}
\begin{figure}[!t]
\centering\includegraphics[width=0.9\linewidth]{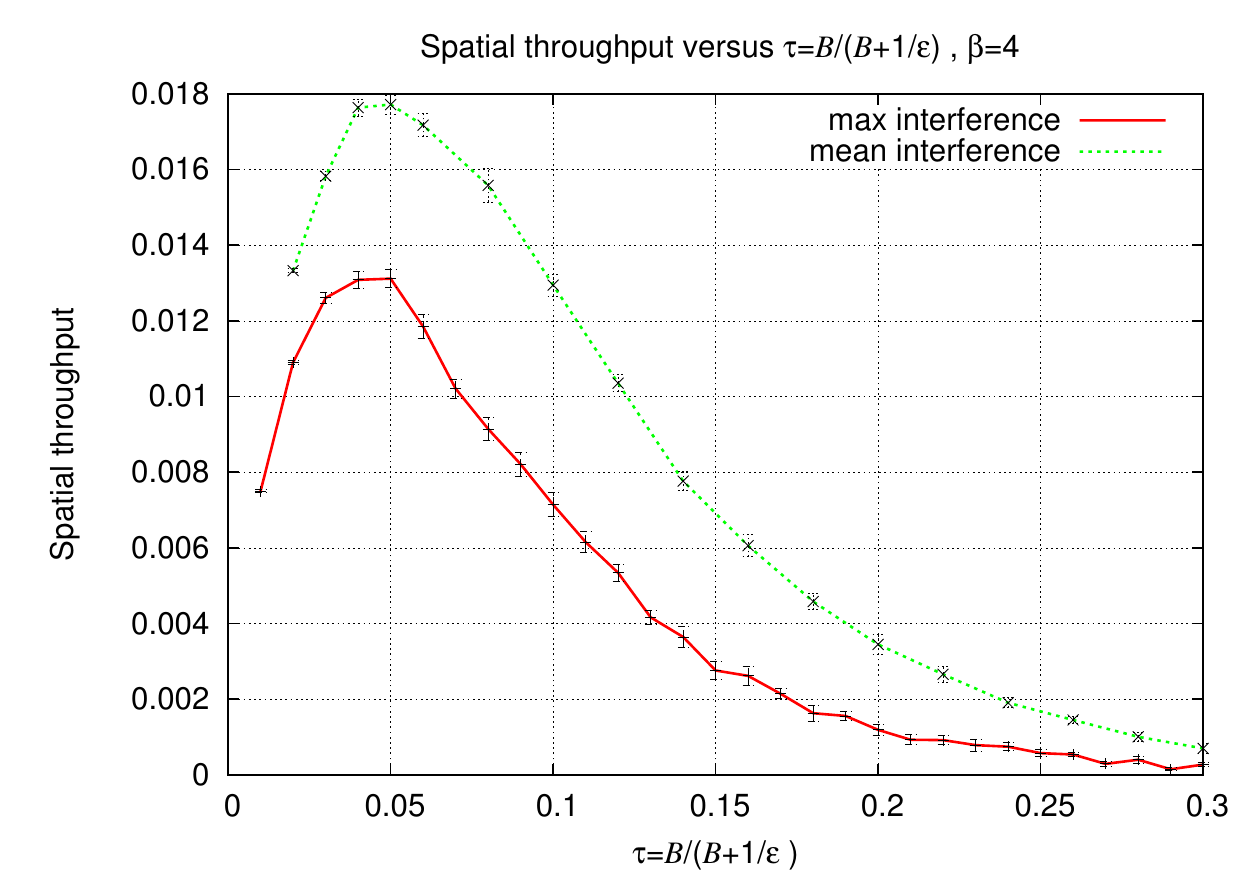}
\caption{Spatial throughput versus
  $\frac{B}{B+1/\epsilon}$ for mean and maximal interference constraint; simulation
  results. }
\label{fig.ns_aloha_mean_max_b41}
\end{figure}

In this section, we show the impact of the 
assumption on the maximum interference
constraint in the SINR
on the probability of a successful transmission.
In Figure~\ref{fig.ns_aloha_mean_max_b41}, we compare the spatial 
throughput for non slotted Aloha, we still have 
$\lambda = 1$, $r=1$, $T=10$ and $\beta=4$.
When the SINR is computed with the maximum interference instead of 
the mean interference the loss in performance 
is large and may be up to $45\%$. But if the 
throughput is optimized in $\epsilon$, the loss in performance is only  
$26\%$. We observe that the throughput is optimized in both 
cases for the same value of $B\epsilon \simeq 0.045$, this value 
is also optimal for the Poison rain model.

%\toself Comparison of $p^{max}_{ren}$ to $p_{ren}^{mean}$; TODO

%% In Figure~\ref{fig.ns_aloha_mean_max_b4} we compare $p^{max}_{ren}$ to 
%% $p_{ren}^{mean}$  for   $\lambda = 0.001$, $r=\sqrt{1000}$, 
%% $T=10$ and $\beta=4$.
%% The loss in performance when the SINR is computed with the maximum 
%% interference can be large and may be up to $45\%$. But when the 
%% throughput is optimized in $\epsilon$, the loss in performance is only  
%% $26\%$. We observe that the throughput is optimized in both 
%% cases for the same value of $B\epsilon \simeq 0.045$, this value 
%% is also optimal for the Poison rain model.
%% The density of successful transmissions for 
%% non-slotted Aloha when the SINR is not averaged is  $55\%$ 
%% that of slotted Aloha. In this case, the comparison is  close to 
%% those of the 'standard' model of slotted/non-slotted Aloha on a 
%% wired network.

\begin{figure}[!t]
\centering\includegraphics[width=0.9\linewidth]{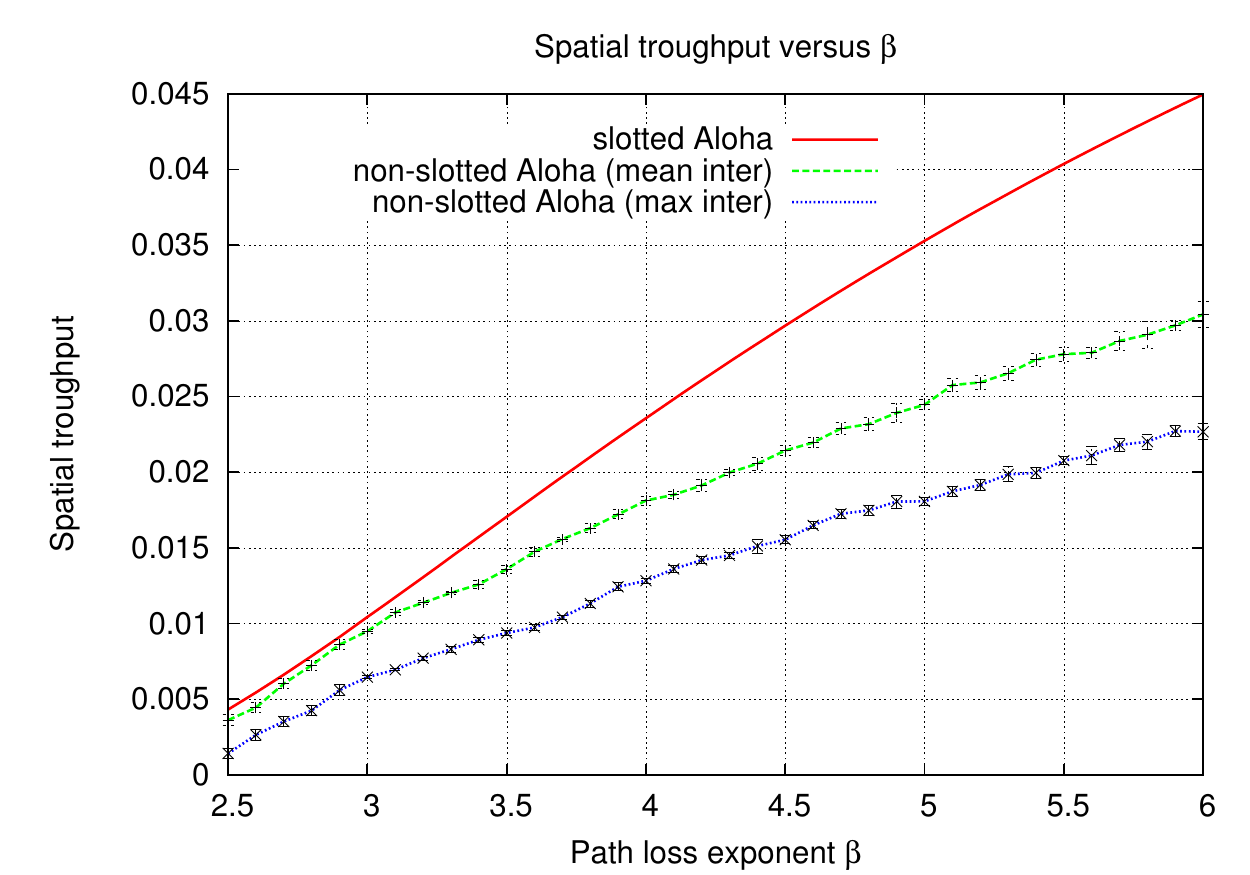}
\caption{Spatial throughput versus path loss exponent
 $\beta$. Slotted Aloha and non-slotted Aloha (mean and maximal interference constraint) are tuned to maximize the spatial throughput. }
\label{fig.ns_aloha_mean_max_bvar}
\end{figure}

In Figure~\ref{fig.ns_aloha_mean_max_bvar} we compare the spatial 
throughput for slotted Aloha and non-slotted Aloha when 
the maximum or average SINR is considered. For slotted Aloha, we use 
the analytical model and optimize spatial throughput in $p$. 
For non-slotted Aloha, we use simulation results and the Poison rain 
model to optimize the schemes in $\frac{B}{B+1/\epsilon}=\tau$. We observe 
that for, $\beta \leq 4$  non-slotted Aloha with the averaged SINR 
provides $50\%$ more throughput than with the maximum SINR.  
For $\beta \geq 5$, non-slotted Aloha with the averaged SINR 
provides only around $35\%$ more throughput than with the maximum SINR.
When we compare slotted Aloha with non-slotted Aloha with maximum 
SINR, we find that slotted Aloha offers $66\%$ more throughput for 
$\beta=3$ and $100\%$ for $\beta=6$.

\section{Conclusion}
\label{sec:con}
Following on from previous studies of slotted Aloha, 
we propose two models for pure Aloha: the Poisson rain and the Poisson 
renewal models.  We provide simulation results to prove the
validity of both models. We show that the Poisson renewal model 
converges towards the Poisson rain model when the density of nodes 
becomes large. As the main theoretical contribution we derive 
closed form expressions  for the Laplace transform of the interference
observed in  both models with a general fading distribution.
For the Poisson Rain model these expressions are very simple
and particularly amenable to comparison with the previously well studied
slotted Aloha model.  Using this approach 
we express the gain in spatial throughput for slotted Aloha (due to synchronisation) 
as a simple function  of the path-loss exponent. It indicates that
for small values of the path-loss exponent  (close to 2) there is no
gain from synchronisation, and that this gain very slowly increases 
with the increase in the path loss exponent (theoretically up to the factor of 2,
when the path-loss exponent approaches infinity).

\appendix

\subsection{Proof of Proposition~\ref{prop_li_g}}
\label{appendix}
Each node  $X_j$, $j\not=0$, sends only  two packets
which can potentially interfere with the transmission of the packet of
node $X_0=0$, which starts at time $T_0(0)=0$. These are 
the last transmission (of  $X_j$) which starts
before time~0 (at time $T_j(0)$) and the first one which starts after
time~0 (at time  $T_j(1)$   with 
$T_j(0)\le 0<T_j(1)$) . With this notation we have~\footnote{We have: $h(s)=(B-|s|)^+/B$, see Proposition~\ref{fact.nonslotted_kappa}  .} : 
\begin{align}\label{e.I_renewal}
& I^{mean} = \\
& \sum_{X_j\in\Phi}F_j^0(0)h(T_j(0))/l(|X_j|) + F_{j}^0(1)h(T_j(1))/l(|X_j|)\nonumber  \,.
\end{align}
where $F_j^0(0)$ and $F_j^0(1)$ are the fading variables between station $X_j$ and $X_0$ during the 
transmissions of the two potentially interfering packets described above. 
To simplify let us denote $F_j(0)=F_j^0(0)$,
$F_j(1)=F_j^0(1)$, $H_j(0)=h(T_j(0))$, $H_j(1)=h(T_j(1))$. 
%
%$$I^{mean} = \sum_{X_j\in\Phi} F_j(n)/l(|X_j|)h(T_j(n)) $$
%$$I^{mean} = \sum_{X_j\in\Phi} \sum_{n=0,1} F_j(n)/l(|X_j|)h(T_j(n)) $$
%$$H_j(0)=h(T_j(0))$$ 
%$$H_j(1)=h(T_j(1))$$ 
Let us consider the marked point process
$$\tilde{\Phi}=\{X_j,\,\Bigl((F_j(0),F_j(1),H_j(0),H_j(1)\Bigr): j\not=0\}\,.$$
It is an independently marked Poisson process with points
$\{X_j:j\not=0\}$, and marks $\{((F_j(0),F_j(1),H(0),H(1))\}$
(independent across $j$, given points). Note that for given $j$, $F_j(0)$, $F_j(1)$
and the vector $(H_j(0),H_j(1))$ are mutually independent, however, $H_j(0)$
and $H_j(1)$ are not independent from each other (we 
describe their joint distribution below). Let us consider the following
mapping of $\tilde{\Phi}$
$$\tilde{\Psi}=\{(F_j(0)/l(X_j))^{-1},\,\Bigl((F_j(1)/l(X_j))^{-1},H_j(0),H_j(1)\Bigr)\}$$
considered again as a marked point process, with points
$\{V_j:=F_j(0)/l(X_j))^{-1}\}$
and marks  $\Bigl((F_j(1)/l(X_j))^{-1},H_j(0),\allowbreak H_j(1)\Bigr)$.
By the displacement theorem it is again an independently marked
Poisson point process, with intensity (of points)
$\Lambda(0,s)=as^{2/\beta }$ where $a=\frac{\lambda \pi
  \E[F^{\frac{2}{\beta}}]}{A^2}$. Regarding its marks,
$(H_j(0),H_j(1))$ are identically distributed vectors (as in
$\tilde{\Phi}$). However, $(F_j(1)/l(X_j))^{-1}$, being independent of
$(H_j(0),H_j(1))$,  has a distribution which depends on the value of $V_j$. 
This distribution can be represented as follows: the law of
$(F_j(1)/l(X_j))^{-1}$ given 
$V_j=v$ is equal to the law of $l(R_v)/F$ where 
$$\Pro(R_v \leq r)= \frac{1}{\E[F^{\frac{2}{\beta}}]} \E\big[\ind(F \leq
\frac{l(r)}{v}\big) F^{\frac{2}{\beta}}]\,,$$
with $F$ having the distribution of fading (hence the same as $F_j(0)$
and $F_j(1)$). Using these observations and the well known formula
for the Laplace transform of the Poisson point process we obtain
%$$I = \sum_{V_i\in\Psi} V_i^{-1}H_i(0)+ \frac{F_i}{l(R_i)}H_i(1)  $$
%
\begin{align*}
\calL_{I^{mean}}(\xi)&= \E\big[e^{-\xi I}\big]\\
&= \exp\big(-\int_0^{\infty} (1-
\E\big[\frac{H(0)}{v}+\frac{H(1)F}{l(R_v)}\big] )\Lambda(\md v)
\big)\,.
\end{align*}
We focus now on the joint distribution of $(H(0),H(1))$~\footnote{we simplify the 
notation $H(0)=H_j(0)$, $H(1)=H_j(1))$}.
Let $U[x,y]$ be the uniform law on $[x,y]$ and $\epsilon_0,\epsilon_1$ two 
independent exponential variables of rate $\epsilon$. According to the 
renewal theory (see e.g.~\cite[eq.~1.4.3]{BaccelliBremaud2003}),  
 we have the following result. $T(0)= U[-B,0]$ 
with probability $\frac{\epsilon B}{1+\epsilon B}$ and $T(0)=-(B+\epsilon_0)$
with probability $\frac{1}{1+\epsilon B}$. 
$T(1)= B+T(0)+\epsilon_1$ if $T(0)>  -B$ and $T(1)= \epsilon_1$ otherwise. 
Thus we have $H(0)=h(-U)$ and $H(1)=h(-U+B+\epsilon_1)$ with probability 
$\frac{\epsilon B}{1+\epsilon B}$  and 
$H(0)=h(-U-\epsilon_0 )$ and $H(1)=h(\epsilon_1)$ 
with probability $\frac{1}{1+\epsilon B}$, where $U$ is $U[0,B]$.
Consequently, 
\begin{align*}\label{L_I}
%%\lefteqn{\calL_I}\\
&\E\big[\E[e^{-\xi(\frac{H(0)}{v}+\frac{H(1)F}{l(R_v)}|H(0),H(1))})]\big]\\
&= \E\big[e^{-\xi \frac{H(0)}{v}} \calL_{F/lR_v}(\xi H(1))\big] \\
&= \;\;\frac{\epsilon B}{1+\epsilon B} \E\big[e^{-\xi \frac{h(-U)}{v}}
\calL_{F/lR_v}(\xi h(-U+B+\epsilon_1))\big] \\
&\;\;+  \frac{1}{1+\epsilon B} \E\big[e^{-\xi \frac{h(-B-\epsilon_0 )}{v}} \calL_{F/lR_v}(\xi h(\epsilon_1))\big] \,  
\end{align*} 
where
\begin{eqnarray*}\label{L_v}
\calL_{F/lR_v}(\xi) & = &   \E\big[\E[e^{-\xi F/l(R_v)}]\big] \\
             & = &   \E\big[F'^{2/\beta} e^{-\xi F/l(vF')^{1/\beta}}\big]\,.
\end{eqnarray*}
with $F'$ being independent of $F$ with the same distribution. 
Note that $\calL_{F/lR_v}(\xi)=\tilde\calL(\xi/v)$, where
$\tilde\calL(\xi)=\E\big[F'^{2/\beta} \calL_F(\xi /F')\big]$.
Thus, we have: 
\begin{align*}%\label{L_v}
&\E\big[\E[e^{-\xi(\frac{H(0)}{v}+\frac{H(1)F}{l(R_v)}|H(0),H(1))}]\big]\\
&=    \frac{\epsilon B}{(1+\epsilon B)B} \int_0^B e^{-\xi h(-u)/v} \E[\tilde{\calL}(\xi h(-u+B+\epsilon_1)/v] \md u \\
             &\hspace{0.2\linewidth} +   \frac{1}{1+\epsilon B} \E[e^{-\xi h( (-B-\epsilon_0)/v)}]  \E[\tilde{\calL}(\xi h(\epsilon_1)/v] \\
             & =    \frac{\epsilon B}{(1+\epsilon B)B} \int_0^B e^{-\xi h(-u)/v} \E[\tilde{\calL}(\xi h(-u+B+\epsilon_1)/v] \md u \\
&\hspace{0.2\linewidth} +   \frac{1}{1+\epsilon B} \E[\tilde{\calL}(\xi h(\epsilon_1)/v].
\end{align*}
 We set : 
\begin{align*}
E_1&= \E[\tilde{\calL}(\xi h(-u+B+\epsilon_1)/v] \\
E_2&= \E[\tilde{\calL}(\xi h(\epsilon_1)/v]\\
F_1&=\frac{1}{B} \int_0^B e^{-\xi h(-u)/v} E_1 \md u.
\end{align*}
Let us denote $\eta = \xi /v$ and calculate:
\begin{align*}
E_1&= \epsilon \int_0^u e^{-\epsilon s} \tilde{\calL}\Bigl(\frac{\eta (u-s)}{ B}\Bigr) \md s + e^{-\epsilon u}\\
E_2&= \epsilon \int_0^B  e^{-\epsilon s} \tilde{\calL}\Bigl (\frac{\eta (B-s)}{ B}\Bigr) \md s + e^{-\epsilon B}\\ 
\end{align*}
and 
\begin{align*}
F_1&=\frac{\epsilon}{B} \int_0^B e^{-\eta (1-u/B)} \int_0^u
e^{-\epsilon s} \tilde{\calL}\Bigl(\eta\frac{u-s}{B}\Bigr) \md s \md u
\\
&\hspace{0.1\linewidth}
+ \frac{1}{B} \int_0^B e^{-\eta(1-u/B)} e^{-\epsilon u}\md u\,.
\end{align*}
Using the change of variable $\frac{u-s}{B}=t$ we obtain: 
\begin{align*}
F_1&= \int_0^1 \frac{\epsilon B}{\eta- \epsilon
  B}\bigl(e^{\eta-\epsilon B}-e^{t\eta-t\epsilon B }\bigr) \calL(\eta
t) \md t\\
&+\;\; \frac{e^{-\eta}}{\epsilon B -\eta }(1-e^{\eta-\epsilon
  B})\\
&= \frac{\epsilon B e^{\eta-\epsilon B}}{\eta- \epsilon B} \int_0^1
e^{-\epsilon B t}  (1-e^{(\eta-\epsilon B)(t-1) }) \calL(\eta t) \md t\\
&+\;\; \frac{e^{-\eta}}{\epsilon B -\eta }(1-e^{\eta-\epsilon B}).\\
\end{align*}
Denoting
\begin{eqnarray*}
F_2=E_2&=& \epsilon B e^{-\epsilon B} \int_0^1 e^{\epsilon tB}\calL(\eta t) \md t + e^{-\epsilon B}\\
\end{eqnarray*}
we have: 
\begin{align*}
&\calL_{I^{mean}}(\xi)\\
&= \exp \Big[-\int_0^{\infty}  \big(1-\frac{\epsilon B}{(1+\epsilon B)}  F_1 - \frac{1}{(1+\epsilon B)} F_2   \big) \Lambda(dv) \Big]
\end{align*}
which gives the result presented. 
%\end{proof}

\singlespacing
\pdfbookmark[0]{References}{References}
\bibliographystyle{plainnat}
%\bibliography{ieeeit_sns}

\end{document}